# Full-field Fourier ptychography (FFP): spatially varying pupil modeling and its application for rapid field-dependent aberration metrology

Pengming Song,[1,a)] Shaowei Jiang,[2,a)] He Zhang,[2] Xizhi Huang,[2,3] Yongbing Zhang,[3] and Guoan Zheng[1,2,b)]

[1]*Department of Electrical and Computer Engineering, University of Connecticut, Storrs, Connecticut, 06269, USA*

[2]*Department of Biomedical Engineering, University of Connecticut, Storrs, Connecticut, 06269, USA*

[3]*Shenzhen Key Lab of Broadband Network and Multimedia, Graduate School at Shenzhen, Tsinghua University, Shenzhen, Guangdong, 518055, China*

**Abstract**: Digital aberration measurement and removal play a prominent role in computational imaging platforms aimed at achieving simple and compact optical arrangements. A recent important class of such platforms is Fourier ptychography, which is geared towards efficiently creating gigapixel images with high resolution and large field of view (FOV). In current FP implementations, pupil aberration is often recovered at each small segment of the entire FOV. This reconstruction strategy fails to consider the field-dependent nature of the optical pupil. Given the power series expansion of the wavefront aberration, the spatially varying pupil can be fully characterized by tens of coefficients over the entire FOV. With this observation, we report a Full-field Fourier Ptychography (FFP) scheme for rapid and robust aberration metrology. The meaning of 'full-field' in FFP is referred to the recovering of the 'full-field' coefficients that govern the field-dependent pupil over the entire FOV. The optimization degrees of freedom are at least two orders of magnitude lower than the previous implementations. We show that the image acquisition process of FFP can be completed in ~1s and the spatially varying aberration of the entire FOV can be recovered in ~35s using a CPU. The reported approach may facilitate the further development of Fourier ptychography. Since no moving part or calibration target is needed in this approach, it may find important applications in aberration metrology. The derivation of the full-field coefficients and its extension for Zernike modes also provide a general tool for analyzing spatially varying aberrations in computational imaging systems.

## 1. Introduction

Characterization of spatially varying wavefront aberration is of critical importance in ophthalmology, microscopy, astronomy, photography, and lithography. The knowledge of system aberration allows aberration correction either actively through adaptive optics or passively with post-acquisition aberration removal. Different aberration characterization methods have been reported in past years, including interferometry[1, 2], wavefront sensing[3-5], phase retrieval[6, 7], optimization-based spatially-varying pupils recovery[8], point spread function measurement[9], calibration masks[10-12], statistical modeling of weakly scattering object[13], differential phase contrast imaging[14], and etc. In ptychography, spatially varying probe beam can be recovered using a orthogonal relaxation approach[15]. Among these different implementations, digital aberration measurement and removal play an especially prominent role in computational imaging platforms aimed at achieving simple and compact optical arrangements. A recent important class of such platforms is Fourier ptychography[16-27], which is geared towards efficiently creating gigapixel images with high resolution over a large field-of-view (FOV). In a standard FP system, we illuminate the object with angle-varied plane waves and collect the corresponding images through a low numerical aperture (NA)

---

a) P. Song and S. Jiang contributed equally to this work
b) Author to whom correspondence should be addressed. Email: guoan.zheng@uconn.edu

objective lens. As such, each captured image represents the information of a circular disk in the Fourier domain and its offset from the origin is determined by the illumination angle. We can then apply the phase retrieval process to synthesize the aperture information in the Fourier domain. Unlike the conventional synthetic aperture imaging approach, FP requires no phase information to stitch the apertures in the Fourier domain. Instead, the phase information is recovered thanks to the partial aperture overlap between adjacent acquisitions[28, 29]. At the end of the FP reconstruction process, the synthesized information in the Fourier domain generates a higher resolution complex object image that retains the original large FOV set by the low-NA objective.

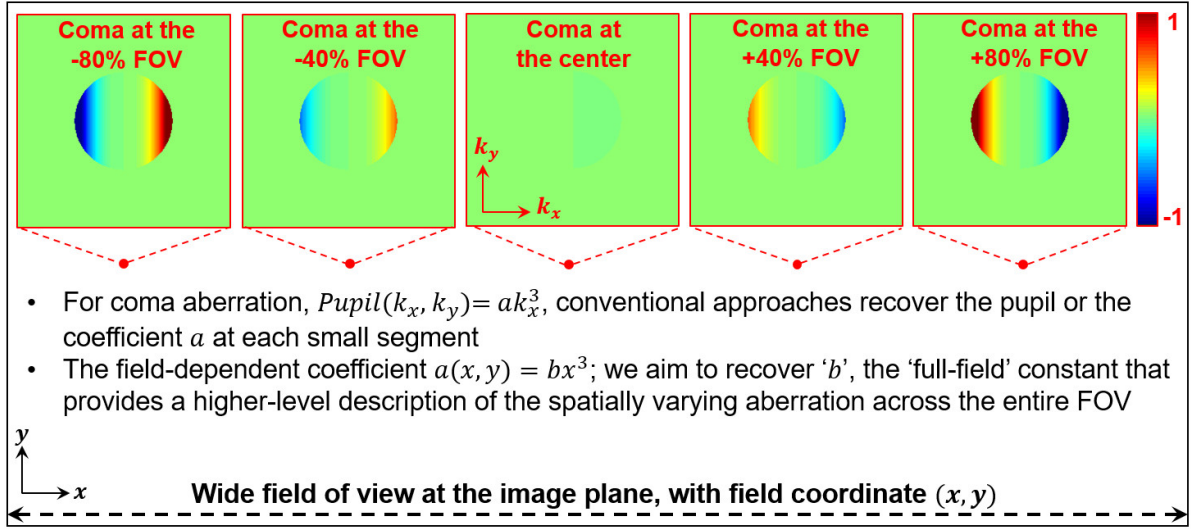

Fig. 1 We use $(k_x, k_y)$ to denote the pupil coordinate at the Fourier plane and $(x, y)$ to denote the field coordinate at the image plane. The pupil aberration $Pupil(k_x, k_y)$ is varying at different field coordinate $(x, y)$. Here we use off-axis coma aberration as an example with $pupil(k_x, k_y) = a \cdot k_x^3$, where '$a$' represents the mount of coma aberration. As one expects, the amount of coma is 0 at the center and become larger at the edge of the FOV. The field-dependent coefficient $a(x, y)$ follows the form of $bx^3$ and we aim to recover the 'full-field' constant '$b$' via Fourier ptychography in this work. Such full-field constants provide a higher-level description of the spatially varying aberration across the entire FOV.

In current FP implementations, we need to divide the entire FOV into smaller segments to handle the spatially varying pupil aberrations. At each small segment, the pupil aberration is treated as spatially invariant and can be locally recovered in an optimization process[14, 19, 30, 31]. Such an aberration recovery strategy, however, fails to consider the field-dependent nature of the optical pupil. Given the power series expansion of the wavefront aberration, the spatially varying pupil can be fully characterized using tens of power series coefficients over the entire FOV. As an example, we show the field-dependent coma aberration in Fig. 1, where we use $(x, y)$ to represent the field location of the image plane and $(k_x, k_y)$ to represent the wavevector location of the pupil (Fourier) plane. For the coma aberration shown in Fig. 1, the wavefront follows the form of $k_x^3$ in the pupil plane and the coefficient $a(x, y)$ represents the amount of coma aberration at the field location $(x, y)$. To the best of our knowledge, most existing aberration recovery schemes aim to recover the pupil or the coefficient $a(x, y)$ for a given field location $(x, y)$. As such, all different segments of the FOV are treated independently and the recovered coefficient $a$ is only related to the local segment. As we will discuss in the next section, the coefficient $a(x, y)$ in Fig. 1 follows the form of $bx^3$ in the image plane, where $b$ is a power series constant across the entire FOV. Therefore, it is better to recover such a 'full-field' constant '$b$' that provides a higher-level description of the spatially varying aberration across the entire FOV.

In this work, we report a Full-field Fourier Ptychography (FFP) scheme for rapid and robust aberration metrology. The meaning of 'full-field' in FFP is referred to the recovering of the 'full-field' power series constants that govern the field-dependent pupil over the entire FOV. We show that the image acquisition process of FFP can be completed in ~1s and the spatially varying aberration of the entire FOV can be recovered in ~35s using a regular CPU. We also demonstrate the reported approach for generating a gigapixel image of a biological sample with a ~8-mm FOV and a 0.5 synthetic NA.

Our approach has several advantages over the existing techniques for aberration metrology. First, we recover the 'full-field' constants instead of the pupil wavefront at each small segment. As such, the optimization degrees of freedom are orders of magnitude lower than the previous implementations. The field-dependent power series expansion can also be viewed as a strong constraint for the pupil recovery process, improving the robustness and accelerating the convergence speed. Second, the full-field constants are jointly recovered using segments over the entire FOV. As such, our reconstruction can be viewed as an averaging result across the entire FOV. It is, in general, more accurate than the localized recovery in a regular implementation. Third, we require no mechanical scanning in our approach. The only hardware modification is to place a low-cost LED array under the object. Such a simple and low-cost configuration enables aberration metrology in systems where traditional techniques are difficult to apply. Fourth, no pre-known calibration mask is needed in our approach. In the reconstruction process, we jointly recover the pupil aberration and the unknown object simultaneously. We use a low-cost blood smear slide in our demonstrations and any thin section would work in our approach.

This paper is structured as follows. In Section 2, we will discuss the spatially varying pupil modeling and derive the field-dependent expression for both power series polynomials and Zernike modes. In Section 3, we will discuss the forward imaging model and the reconstruction process of the FFP approach. In Section 4, we will present our experimental measurements using the FFP approach and quantify the measurement errors. Finally, we will summarize the results in Section 5. The reported approach may greatly facilitate the development of Fourier ptychographic imaging platforms. Since no moving part or calibration target is needed in this approach, it may also find important applications in aberration metrology. The derivation of the full-field power series coefficients and its extension for Zernike modes also provide a general tool for analyzing spatially varying aberrations in computational imaging systems.

## 2. Spatially varying pupil aberration modeling and full-field coefficients

It is well-known that the pupil aberration depends on the field location[32]. In this section, we will derive the field-dependent expression via power series expansions and then extend it for Zernike modes. As shown in Fig. 2, we consider the light wave propagates from the pupil plane with coordinates $(k_x, k_y)$ to the image plane with coordinates $(x, y)$. If we fix the field location to be $(x_0, y_0)$, the wavefront aberration will be a function of pupil coordinates only, i.e., $W(k_x, k_y)$, and it represents the regular treatment of spatially-invariant aberration. To obtain the complete information about the system aberrations, we need to consider wavefront aberration at different field locations and it becomes a function of 4 variables $W(k_x, k_y, x, y)$. These 4 variables are needed for optical systems without rotational symmetry. With rotational symmetry assumption, we can simplify it to $W(k_x^2 + k_y^2, xk_x + yk_y, x^2 + y^2)$, where three variables are enough to describe the system aberrations[32].

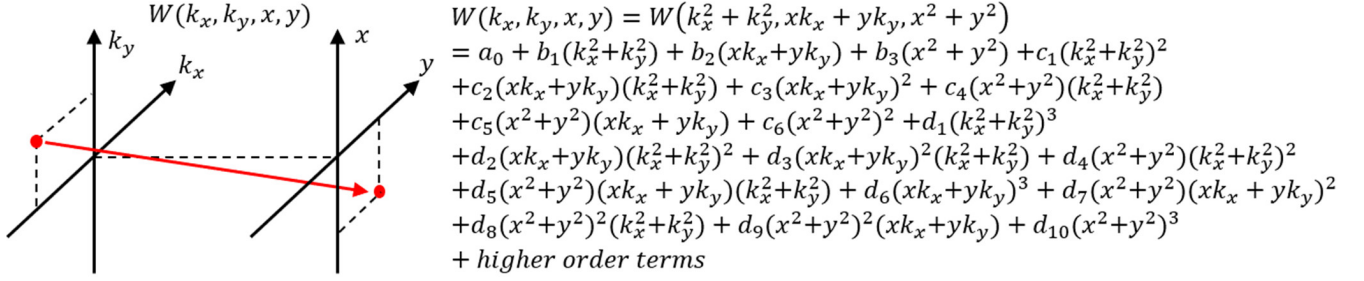

$$W(k_x, k_y, x, y) = W(k_x^2 + k_y^2, xk_x + yk_y, x^2 + y^2)$$
$$= a_0 + b_1(k_x^2+k_y^2) + b_2(xk_x+yk_y) + b_3(x^2 + y^2) + c_1(k_x^2+k_y^2)^2$$
$$+ c_2(xk_x+yk_y)(k_x^2+k_y^2) + c_3(xk_x+yk_y)^2 + c_4(x^2+y^2)(k_x^2+k_y^2)$$
$$+ c_5(x^2+y^2)(xk_x + yk_y) + c_6(x^2+y^2)^2 + d_1(k_x^2+k_y^2)^3$$
$$+ d_2(xk_x+yk_y)(k_x^2+k_y^2)^2 + d_3(xk_x+yk_y)^2(k_x^2+k_y^2) + d_4(x^2+y^2)(k_x^2+k_y^2)^2$$
$$+ d_5(x^2+y^2)(xk_x + yk_y)(k_x^2+k_y^2) + d_6(xk_x+yk_y)^3 + d_7(x^2+y^2)(xk_x + yk_y)^2$$
$$+ d_8(x^2+y^2)^2(k_x^2+k_y^2) + d_9(x^2+y^2)^2(xk_x+yk_y) + d_{10}(x^2+y^2)^3$$
$$+ higher\ order\ terms$$

Fig. 2 Spatially varying pupil aberration modeling. The wavefront aberration is expanded to a power series, with field-dependence on each term.

In Fig. 2, we expand this aberration expression as a power series in three variables. Based on the power series expansion, we summarize the aberration expressions and its field-dependent function in Fig. 3. The first term, $k_x^2 + k_y^2$, represents the defocus aberration and its field dependence is a constant plus a quadratic and a biquadratic term. The second term in Fig. 3, $k_x^2$, represents the astigmatism aberration and its field dependent function has a quadratic and a biquadratic term. Similarly, $k_x^3$ represents coma aberration and its field dependence is a cubic function. We note that, the field-dependent function in Fig. 3 also provide some physical insights on the aberrations. For example, coma aberration is often known as an off-axis aberration and it will vanish at the on-axis position. This point can be verified by the field-dependent function $d_6 x^3$ which equals to 0 at the on-axis position. In our FFP recovery scheme, we aim to recover the power series coefficients, $b_1$, $c_4$, $d_8$, etc. We term these coefficients as 'full-field coefficients', which govern the spatially varying aberrations over the entire FOV. Compare to the recovery of the aberrations of a local segment, the full-field coefficients represent a higher-level description of system aberrations. Tens of such coefficients can fully characterize the aberration over the entire FOV. As such, the optimization degrees of freedom are at least two orders of magnitude lower than the previous approaches, enabling a rapid and robust aberration metrology scheme.

| Pupil plane | Field-dependent function | Pupil plane | Field-dependent function |
|---|---|---|---|
| $k_x^2 + k_y^2$ | $b_1 + c_4(x^2+y^2) + d_8(x^2+y^2)^2$ | $k_x k_y^2$ | $3d_6 xy^2$ |
| $k_x^2$ | $c_3 x^2 + d_7 x^2(x^2+y^2)$ | $(k_x^2+k_y^2)^2$ | $c_1 + d_4(x^2+y^2)$ |
| $k_y^2$ | $c_3 y^2 + d_7 y^2(x^2+y^2)$ | $k_x^2(k_x^2+k_y^2)$ | $d_3 x^2$ |
| $k_x k_y$ | $2c_3 xy + d_7 xy(x^2+y^2)$ | $k_y^2(k_x^2+k_y^2)$ | $d_3 y^2$ |
| $k_x(k_x^2+k_y^2)$ | $c_2 x + d_5 x(x^2+y^2)$ | $k_x k_y(k_x^2+k_y^2)$ | $2d_3 xy$ |
| $k_y(k_x^2+k_y^2)$ | $c_2 y + d_5 y(x^2+y^2)$ | $k_x(k_x^2+k_y^2)^2$ | $d_2 x$ |
| $k_x^3$ | $d_6 x^3$ | $k_y(k_x^2+k_y^2)^2$ | $d_2 y$ |
| $k_y^3$ | $d_6 y^3$ | $(k_x^2+k_y^2)^3$ | $d_1$ |
| $k_x^2 k_y$ | $3d_6 x^2 y$ | | |

Fig. 3 Pupil-plane wavefront aberrations and their field-dependence. We aim to recover the power series coefficients ($b_1$, $c_4$, $d_8$, ...) in our FFP scheme. These coefficients are termed 'full-field coefficients', which govern the spatially varying aberrations over the entire FOV.

Zernike aberrations are combinations of different Taylor polynomials terms shown in Fig. 3. We also provide the field-dependent functions of Zernike aberrations in Fig. 4, where we ignore the biquadratic terms in the field-dependent function. We also note that, the coefficients in Fig. 4, $a_1, a_2, a_3\ldots$, are not independent with each other. For example, $a_5 = a_7$ and $a_6 = a_8$ for a rotational symmetry system. The derivation of their relationship is beyond the scope of this work.

| Zernike polynomials | Pupil plane | Field-dependent function |
| --- | --- | --- |
| $Z_2^0$ | $2\sqrt{3}(k_x^2 + k_y^2) - \sqrt{3}$ | $a_1(x^2+y^2) + a_2$ |
| $Z_2^2$ | $\sqrt{6}(k_x^2 - k_y^2)$ | $a_3(x^2 - y^2)$ |
| $Z_2^{-2}$ | $2\sqrt{6}k_x k_y$ | $a_4 xy$ |
| $Z_3^1$ | $3\sqrt{8}k_x(k_x^2 + k_y^2) - 2\sqrt{8}k_x$ | $a_5 x(x^2+y^2) + a_6 x$ |
| $Z_3^{-1}$ | $3\sqrt{8}k_y(k_x^2 + k_y^2) - 2\sqrt{8}k_y$ | $a_7 y(x^2+y^2) + a_8 y$ |
| $Z_4^0$ | $6\sqrt{5}(k_x^2 + k_y^2)^2 - 6\sqrt{5}(k_x^2 + k_y^2) + \sqrt{5}$ | $a_9(x^2+y^2) + a_{10}$ |
| $Z_3^3$ | $4\sqrt{8}k_x^3 - 3\sqrt{8}k_x(k_x^2 + k_y^2)$ | $a_{11}x^3 + 3a_{12}x(x^2+y^2) + a_{13}x$ |
| $Z_3^{-3}$ | $3\sqrt{8}k_y(k_x^2 + k_y^2) - 4\sqrt{8}k_y^3$ | $a_{14}y(x^2+y^2) + a_{15}y^3 + a_{16}y$ |
| $Z_4^2$ | $(4\sqrt{10}(k_x^2+k_y^2) - 3\sqrt{10})(k_x^2 - k_y^2)$ | $a_{17}(x^2 - y^2)$ |
| $Z_4^{-2}$ | $8\sqrt{10}k_x k_y(k_x^2 + k_y^2) - 6\sqrt{10}k_x k_y$ | $a_{18}xy$ |
| $Z_5^1$ | $10\sqrt{12}k_x(k_x^2 + k_y^2)^2 - 12\sqrt{12}k_x(k_x^2 + k_y^2) + 3\sqrt{12}k_x$ | $a_{19}x(x^2+y^2) + a_{20}x$ |
| $Z_5^{-1}$ | $10\sqrt{12}k_y(k_x^2 + k_y^2)^2 - 12\sqrt{12}k_y(k_x^2 + k_y^2) + 3\sqrt{12}k_y$ | $a_{21}x(x^2+y^2) + a_{22}x$ |
| $Z_6^0$ | $20\sqrt{7}(k_x^2 + k_y^2)^3 - 30\sqrt{7}(k_x^2 + k_y^2)^2 + 12\sqrt{7}(k_x^2 + k_y^2) - \sqrt{7}$ | $a_{23}(x^2+y^2) + a_{24}$ |
| $Z_8^0$ | $70\sqrt{9}(k_x^2 + k_y^2)^4 - 140\sqrt{9}(k_x^2 + k_y^2)^3 + 90\sqrt{9}(k_x^2 + k_y^2)^2 - 20\sqrt{9}(k_x^2 + k_y^2) + \sqrt{9}$ | $a_{25}(x^2+y^2) + a_{26}$ |

Fig. 4 Zernike modes and their field-dependence.

## 3. Full-field Fourier ptychography (FFP)

### 3.1 Forward imaging model and the optimization problem

In the proposed FFP approach, the forward imaging model is similar to the original FP paper[16]. We use $J$ different angle-varied plane waves to illuminate complex object $o(x, y)$ and acquire the corresponding intensity images $I_j(x, y)$ ($j = 1,2,3\ldots J$) via a low-NA system. In our implementation, we divide the captured image $I_j(x, y)$ into $M$ different small segments and obtain $I_{mj}(x, y)$, where $m = 1,2,3\ldots M$. For each small segment, the wavefront aberration is treated as a spatially invariant pupil and we can obtain the following forward imaging model for the $m^{\text{th}}$ image segment:

$$I_{mj}(x, y) = \left| \left( o_m(x, y) \exp(ik_{xj}x + ik_{yj}y) \right) * psf_m(x, y) \right|^2, \tag{1}$$

where '*' denotes convolution, $o_m(x, y)$ denotes the $m^{\text{th}}$ small segment of the object, and $psf_m(x, y)$ denotes the spatially invariant point spread function (PSF) at the $m^{\text{th}}$ small segment.

The convolution operation in Eq. (1) is typically calculated in the Fourier domain via fast Fourier transform (FFT). We can define the following complex exit wave $\Psi_{mj}(k_x, k_y)$ at the pupil plane using the pupil function $Pupil_m(k_x, k_y)$ as follows:

$$\Psi_{mj}(k_x, k_y) = O_m(k_x - k_{xj}, k_y - k_{yj}) \cdot Pupil_m(k_x, k_y), \tag{2}$$

where $O_m(k_x, k_y)$ is the Fourier transform of $o_m(x, y)$ and $Pupil_m(k_x, k_y)$ is the Fourier transform of $psf_m(x, y)$. Based on the definition in Eq. (2), the forward imaging model of Eq. (1) can be rewritten as

$$I_{mj}(x, y) = |\mathcal{F}^{-1}\{\Psi_{mj}(k_x, k_y)\}|^2 = |\psi_{mj}(x, y)|^2, \tag{3}$$

where $\mathcal{F}^{-1}$ denotes inverse Fourier transform. Assuming the central position of the $m^{th}$ image segment is $(x_m, y_m)$, the pupil function for the $m^{th}$ image segment in Eq. (2) can be expressed as follows

$$Pupil_m(k_x, k_y) = circ\left(NA \cdot \frac{2\pi}{\lambda}\right) \exp\left(i \cdot \sum_{t=1}^{T} f_t(x_m, y_m) \cdot Aber_t(k_x, k_y)\right), \tag{4}$$

where '$circ$' denotes a circular mask with a radius of $NA \cdot \frac{2\pi}{\lambda}$ and $\lambda$ is the wavelength of the light wave. In Eq. (4), the wavefront aberration is represented by $\sum_{t=1}^{T} f_t(x_m, y_m) \cdot Aber_t(k_x, k_y)$, where $f_t(x_m, y_m)$ is the full-field function evaluated at the $(x_m, y_m)$ position and $Aber_t(k_x, k_y)$ is the corresponding aberration mode. We use 17 aberration modes in our implementation, i.e., $T = 17$ in Eq. (4). These 17 aberration modes and their full-field functions are listed in Fig. 3. For example, $Aber_1(k_x, k_y) = k_x^2 + k_y^2$ and $f_1(x_m, y_m) = b_1 + c_4(x_m^2 + y_m^2) + d_8(x_m^2 + y_m^2)^2$; $Aber_2(k_x, k_y) = k_x^2$ and $f_2(x_m, y_m) = c_3 x_m^2 + d_7 x_m^2(x_m^2 + y_m^2)$. Similarly, $Aber_{17}(k_x, k_y) = (k_x^2 + k_y^2)^3$ and $f_{17}(x_m, y_m) = d_1$.

The goal of the FFP approach is to recover the 13 full-field coefficients $\{b_1, c_1, ..., d_8\}$ in 17 field-dependent functions in Fig. 3 (several full-field functions share the same full-field coefficients). We define the following cost function $L_{mj}$ for the $m^{th}$ image segment and the $j^{th}$ incident angle:

$$L_{mj} = \sum_{x,y} \left||\psi_{mj}(x, y)| - \sqrt{I_{mj}(x, y)}\right|^2 \tag{5}$$

The full-field coefficients can be recovered by solving the following optimization problem:

$$\{b_1, c_1, ..., d_8\} = \underset{\{b_1, c_1, ..., d_8, O_m\}}{\arg\min} \sum_{j=1}^{J} \sum_{m=1}^{M} L_{mj} \tag{6}$$

Compared with other pupil recovery schemes, the key innovation here is to impose the constraints of the full-field functions. As such, we only need to recover 13 parameters over the entire FOV. The optimization degrees of freedom are even lower than the spatially invariant blind deconvolution of a single image segment. In our implementation, we divide the entire FOV into ~320 smaller segments and each segment has 256 by 256 pixels. Table 1 lists the optimization degrees of freedom for comparison, where we assume a 64 by 64 kernel size for the blind deconvolution at one image segment.

|  | Update the pupil at 320 segments | Update the 17 aberration modes at 320 segments | Blind deconvolution at one segment | Our full-field model |
|---|---|---|---|---|
| **Optimization degrees of freedom** | 256*256*320 = 20,971,520 | 17*320 = 5440 | 64*64 = 4096 | 13 |

Table 1: Comparison of the optimization degrees of freedom for different approaches.

## 3.2 Recovering full-field coefficients via gradient descent

We provide two solutions for solving the optimization problem in Eq. (6). The first solution is an extension of the EPRY (or ePIE) approach[19, 33]. In the recovery process, we update the object $o(x,y)$ via the EPRY approach and then update the full-field coefficients $\{b_1, c_1, \dots, d_8\}$ via gradient descent. Different from previous single-segment implementations, the gradient update for the full-field coefficients is based on the average from $M$ segments over the entire FOV.

---

**Algorithm 1: Recovering full-field coefficients via gradient descent**

**Definition:** The subscript '$m$' donates different image segments over the entire FOV; the subscript '$j$' donates the $j^{th}$ incident angle for the object; The subscript '$t$' donates the index of the aberration modes

**The loss function:** $L_{mj} = \sum_{x,y} \left\| \mathcal{F}^{-1}\{O_m(k_x - k_{xj}, k_y - k_{yj}) \times Pupil_m(k_x, k_y)\} - \sqrt{I_{mj}(x,y)} \right\|^2$

**Input:** The images $I_{mj}(x,y), m = 1,2,\dots M, j = 1,2,\dots J$

**Output:** The 13 coefficients $\{b_1, c_1, \dots, d_8\}$ of the field-dependent functions $f_t(x,y)$, where $t = 1,2,\dots 17$

**Initialize** $O_m(k_x, k_y), \{b_1, c_1, \dots, d_8\}$, and $Pupil_m(k_x, k_y) = circ(NA \cdot \frac{2\pi}{\lambda}) \cdot \exp[i \cdot \sum_{t=1}^{17} f_t(x_m, y_m) \cdot Aber_t(k_x, k_y)]$

for $n = 1: N$ (iterations)
    for $j = 1: J$ (different incident angles)
        for $m = 1: M$ (different image segments over the entire field of view)
$$Pupil_m(k_x, k_y) = circ(NA \cdot \tfrac{2\pi}{\lambda}) \cdot \exp[i \cdot \sum_{t=1}^{17} f_t(x_m, y_m) \cdot Aber_t(k_x, k_y)]$$
$$\Psi_{mj}(k_x, k_y) = O_m(k_x - k_{xj}, k_y - k_{yj}) \cdot Pupil_m(k_x, k_y)$$
$$\psi_{mj}(x,y) = \mathcal{F}^{-1}\{\Psi_{mj}(k_x, k_y)\}$$
$$\psi_{mj}^{update}(x,y) = \sqrt{I_{mj}} \cdot \exp\{i \cdot angle[\psi_{mj}(x,y)]\}$$
$$\Psi_{mj}^{update}(k_x, k_y) = \mathcal{F}\{\psi_{mj}^{update}(x,y)\}$$
$$O_m(k_x - k_{xj}, k_y - k_{yj}) = O_m(k_x - k_{xj}, k_y - k_{yj}) + \frac{conj[Pupil_m(k_x, k_y)]}{max\{|Pupil_m(k_x, k_y)|^2\}} \cdot [\Psi_{mj}^{update}(k_x, k_y) - \Psi_{mj}(k_x, k_y)]$$
$$\frac{\partial L_{mj}}{\partial b_1} = -2\sum_{x,y}\left\{\left(1 - \frac{\sqrt{I_{mj}(x,y)}}{|\psi_{mj}(x,y)|}\right) \cdot Imaginary\left[conj\left(\psi_{mj}(x,y)\right) \cdot \mathcal{F}^{-1}\{\Psi_{mj}(k_x, k_y) \cdot \nabla_{b_1} f_1(x_m, y_m) \cdot Aber_1(k_x, k_y)\}\right]\right\}$$
$$\frac{\partial L_{mj}}{\partial c_4} = -2\sum_{x,y}\left\{\left(1 - \frac{\sqrt{I_{mj}(x,y)}}{|\psi_{mj}(x,y)|}\right) \cdot Imaginary\left[conj\left(\psi_{mj}(x,y)\right) \cdot \mathcal{F}^{-1}\{\Psi_{mj}(k_x, k_y) \cdot \nabla_{c_4} f_1(x_m, y_m) \cdot Aber_1(k_x, k_y)\}\right]\right\}$$
        ⋮
        end
    $\Delta b_1 = \left(\sum_{m=1}^{M} \frac{\partial L_{mj}}{\partial b_1}\right)/M, \quad b_1 = b_1 - \alpha \cdot \Delta b_1$
    $\Delta c_4 = \left(\sum_{m=1}^{M} \frac{\partial L_{mj}}{\partial c_4}\right)/M, \quad c_4 = c_4 - \alpha \cdot \Delta c_4$
    ⋮
    end
end

---

Fig. 5 Recovering full-field coefficients via gradient descent.

Figure 5 summarizes the recovery process, where the subscript '$m$' denotes different image segments over the entire FOV, the subscript '$j$' denotes the $j^{th}$ incident angle, and the subscript '$t$' denotes the $t^{th}$ aberration modes in Eq. (4). In this process, we first initialize the object spectrum $O_m(k_x, k_y)$ and the 13 full-field coefficients $\{b_1, c_1, \dots, d_8\}$. For each iteration $n$, the images $I_{mj}(x,y)$ are addressed according to different segments and different incident angles. We perform the Fourier magnitude projection as follows:

$$\psi_{mj}^{update}(x,y) = \sqrt{I_{mj}(x,y)} \cdot exp\{i \cdot angle[\psi_{mj}(x,y)]\} \quad (7)$$

The updated complex object is then used to generate the updated spectrum of the exit wave at the pupil plane: $\Psi_{mj}^{update}(k_x, k_y) = \mathcal{F}\{\psi_{mj}^{update}(x, y)\}$. We then update the object spectrum using the following equation:

$$O_m(k_x - k_{xj}, k_y - k_{yj}) = O_m(k_x - k_{xj}, k_y - k_{yj}) + \frac{conj[Pupil_m(k_x, k_y, x_m, y_m)]}{max\{|Pupil_m(k_x, k_y, x_m, y_m)|^2\}} \cdot [\Psi_{mj}^{update}(k_x, k_y) - \Psi_{mj}(k_x, k_y)] \quad (8)$$

Instead of updating the pupil function as in EPRY, we directly update the full-field coefficients $\{b_1, c_1, \ldots, d_8\}$. The gradient of the loss function with respect to $b_1$, $c_4$ are given by:

$$\frac{\partial L_{mj}}{\partial b_1} = \frac{\partial \sum_{x,y} ||\psi_{mj}(x,y)| - \sqrt{I_{mj}(x,y)}|^2}{\partial b_1}$$

$$= \frac{\partial \sum_{x,y} ||\mathcal{F}^{-1}\{O_m(k_x - k_{xj}, k_y - k_{yj}) \times Pupil_m(k_x, k_y)\}| - \sqrt{I_{mj}(x,y)}|^2}{\partial b_1}$$

$$= -2\sum_{x,y}\left\{\left(1 - \frac{\sqrt{I_{mj}(x,y)}}{|\psi_{mj}(x,y)|}\right) \cdot Imaginary\left[conj\left(\psi_{mj}(x,y)\right) \cdot \mathcal{F}^{-1}\{\Psi_{mj}(k_x, k_y) \cdot \nabla_{b_1} f_1(x_m, y_m) \cdot Aber_1(k_x, k_y)\}\right]\right\}, \quad (9)$$

$$\frac{\partial L_{mj}}{\partial c_4} = \frac{\partial \sum_{x,y} ||\psi_{mj}(x,y)| - \sqrt{I_{mj}(x,y)}|^2}{\partial c_4}$$

$$= \frac{\partial \sum_{x,y} ||\mathcal{F}^{-1}\{O_m(k_x - k_{xj}, k_y - k_{yj}) \times Pupil_m(k_x, k_y)\}| - \sqrt{I_{mj}(x,y)}|^2}{\partial c_4}$$

$$= -2\sum_{x,y}\left\{\left(1 - \frac{\sqrt{I_{mj}(x,y)}}{|\psi_{mj}(x,y)|}\right) \cdot Imaginary\left[conj\left(\psi_{mj}(x,y)\right) \cdot \mathcal{F}^{-1}\{\Psi_{mj}(k_x, k_y) \cdot \nabla_{c_4} f_1(x_m, y_m) \cdot Aber_1(k_x, k_y)\}\right]\right\}, \quad (10)$$

where $\nabla_{b_1} f_1(x_m, y_m)$ and $\nabla_{c_4} f_1(x_m, y_m)$ represent the gradient of $f_1(x_m, y_m)$ with respective to $b_1$ and $c_4$, '$conj$' represents complex conjugate, and '$Imaginary$' means taking the imaginary part. The gradient of the loss function with respect to other coefficients can be obtained in a similar manner. For each full-field coefficient, we calculate the update ($\Delta b_1$, $\Delta c_4$, ...) by taking the gradient average of all $M$ small segments:

$$\Delta b_1 = \frac{\left(\sum_{m=1}^{M} \frac{\partial L_{mj}}{\partial b_1}\right)}{M}, \quad \Delta c_4 = \frac{\left(\sum_{m=1}^{M} \frac{\partial L_{mj}}{\partial c_4}\right)}{M}, \ldots \quad (11)$$

The gradient descent method for updating the full-field coefficients is, thus, given by:

$$b_1 = b_1 - \alpha \cdot \Delta b_1, \quad c_4 = c_4 - \alpha \cdot \Delta c_4, \ldots, \quad (12)$$

where α is the step-size for the updating process. This process is repeated for $n$ iterations until the solution converges. In a typical implementation, we use 10-50 loops. With the full-field coefficients, the pupil function at any location of the entire FOV can be obtained via Eq. (4).

### 3.3 Fast implementation via alternating projection

The second solution for solving the optimization problem in Eq. (6) is based on alternating projection. In this case, we first rewrite the pupil function as

$$Pupil_m(k_x, k_y) = circ(NA \cdot \frac{2\pi}{\lambda}) \cdot \exp[i \cdot \sum_{t=1}^{17} \omega_{tm} \cdot Aber_t(k_x, k_y)], \tag{13}$$

where $\omega_{tm}$ represents the aberration coefficient of the $t^{th}$ aberration mode on the $m^{th}$ image segment. In this scheme, we update the coefficients $\omega_{tm}$ and then fit these coefficients to the full-field functions $f_t(x, y)$.

---

**Algorithm 2: Recovering full-field coefficients via alternating projection**

**Definition:** The subscript '$m$' donates different image segments over the entire FOV; the subscript '$j$' donates the $j^{th}$ incident angle for the object; The subscript '$t$' donates the index of the aberration modes

**The loss function:** $L_{mj} = \sum_{x,y} \left\| \left| \mathcal{F}^{-1}\{O_m(k_x - k_{xj}, k_y - k_{yj}) \times Pupil_m(k_x, k_y)\} \right| - \sqrt{I_{mj}(x,y)} \right\|^2$

**Input:** The images $I_{mj}(x, y), m = 1, 2, \ldots M, j = 1, 2, \ldots J$

**Output:** The 13 coefficients $\{b_1, c_1, \ldots, d_8\}$ of the field-dependent functions $f_t(x, y)$, where $t = 1, 2, \ldots 17$

**Initialize** $O_m(k_x, k_y), \{b_1, c_1, \ldots, d_8\}$, and $Pupil_m(k_x, k_y) = circ(NA \cdot \frac{2\pi}{\lambda}) \cdot \exp[i \cdot \sum_{t=1}^{17} \omega_{tm} \cdot Aber_t(k_x, k_y)]$

**for** $n = 1:N$ (iterations)
    **for** $m = 1:M$ (different image segments over the entire field of view)
        **for** $j = 1:J$ (different incident angles)

$$Pupil_m(k_x, k_y) = circ(NA \cdot \frac{2\pi}{\lambda}) \cdot \exp[i \cdot \sum_{t=1}^{17} \omega_{tm} \cdot Aber_t(k_x, k_y)]$$

$$\Psi_{mj}(k_x, k_y) = O_m(k_x - k_{xj}, k_y - k_{yj}) \cdot Pupil_m(k_x, k_y)$$

$$\psi_{mj}(x, y) = \mathcal{F}^{-1}\{\Psi_{mj}(k_x, k_y)\}$$

$$\psi_{mj}^{update}(x, y) = \sqrt{I_{mj}} \cdot \exp\{i \cdot angle[\psi_{mj}(x, y)]\}$$

$$\Psi_{mj}^{update}(k_x, k_y) = \mathcal{F}\{\psi_{mj}^{update}(x, y)\}$$

$$O_m(k_x - k_{xj}, k_y - k_{yj}) = O_m(k_x - k_{xj}, k_y - k_{yj}) + \frac{conj[Pupil_m(k_x, k_y)]}{max\{|Pupil_m(k_x, k_y)|^2\}} \cdot [\Psi_{mj}^{update}(k_x, k_y) - \Psi_{mj}(k_x, k_y)]$$

           **for** $t = 1:17$ (different aberration modes)

$$\frac{\partial L_{mj}}{\partial \omega_{tm}} = -2 \times \sum_{x,y} \left\{ \left(1 - \frac{\sqrt{I_{mj}(x,y)}}{|\psi_{mj}(x,y)|}\right) \cdot Imaginary\left[conj\left(\psi_{mj}(x, y)\right) \cdot \mathcal{F}^{-1}\{\Psi_{mj}(k_x, k_y) \cdot Aber_t(k_x, k_y)\}\right] \right\}$$

$$\omega_{tm} = \omega_{tm} - \alpha \cdot \frac{\partial L_{mj}}{\partial \omega_{tm}}$$

           **end**
        **end**
    **end**

**Fit the updated aberration coefficients** $\omega_{tm}$ ($m = 1, 2, \ldots M$) **to** $f_t(x, y)$, **namely solving:**

$$\begin{cases} \{b_1, c_4, d_8\} = \underset{b_1, c_4, d_8}{\operatorname{argmin}} \sum_{m=1}^{M} |\omega_{1m} - f_1(x_m, y_m)|^2 \\ \quad\quad\quad\quad\vdots \\ \{d_1\} = \underset{d_1}{\operatorname{argmin}} \sum_{m=1}^{M} |\omega_{17m} - f_{17}(x_m, y_m)|^2 \end{cases}$$

**Update the aberration coefficients via** $\omega_{tm} = f_t(x_m, y_m)$
**end**

Fig. 6 Recovering full-field coefficients via alternating projection.

Figure 6 summarizes the recovery process of this scheme. Based on Eq. (5), the gradient of the loss function with respect to the aberration coefficient $\omega_{tm}$ can be expressed as:

$$\frac{\partial L_{mj}}{\partial \omega_{tm}} = \frac{\partial \sum_{x,y} \left||\psi_{mj}(x,y)| - \sqrt{I_{mj}(x,y)}\right|^2}{\partial \omega_{tm}}$$

$$= \frac{\partial \sum_{x,y} \left||\mathcal{F}^{-1}\{O_m(k_x - k_{xj}, k_y - k_{yj}) \times Pupil_m(k_x, k_y)\}| - \sqrt{I_{mj}(x,y)}\right|^2}{\partial \omega_{tm}}$$

$$= -2\sum_{x,y}\left\{\left(1 - \frac{\sqrt{I_{mj}}}{|\psi_{mj}(x,y)|}\right) \cdot Imaginary\left[conj\left(\psi_{mj}(x,y)\right) \cdot \mathcal{F}^{-1}\{\Psi_{mj}(k_x, k_y) \cdot Aber_t(k_x, k_y)\}\right]\right\}. \quad (14)$$

Based on the gradient in Eq. (14), we then update the aberration coefficient $\omega_{tm}$ via

$$\omega_{tm} = \omega_{tm} - \alpha \cdot \frac{\partial L_{mj}}{\partial \omega_{tm}}. \quad (15)$$

With the updated $\omega_{tm}$, we fit it to the full-field functions $f_t(x, y)$ to update the full-field coefficients. For example, given $t = 1$, $\omega_{1m}$ ($m = 1,2,...M$) are fitted to $f_1(x_m, y_m) = b_1 + c_4(x_m^2 + y_m^2) + d_8(x_m^2 + y_m^2)^2$. The fitting process can be considered as the following optimization problem:

$$\{b_1, c_4, d_8\} = \underset{b_1, c_4, d_8}{\operatorname{argmin}} \sum_{m=1}^{M} |\omega_{1m} - f_1(x_m, y_m)|^2 \quad (16)$$

Based on Eq. (16), we can get the full-field functions with the updated full-field coefficients $b_1, c_4, d_8$. We then update the aberration coefficient $\omega_{1m}$ again by projecting it to the full-field functions, i.e., $\omega_{1m} = f_1(x_m, y_m)$. This process is repeated for *n* iterations until the solution converges. The full-field functions can be viewed as constraints for the optimization problem. The updating process of $\omega_{tm}$ can be viewed as a projection operation. This scheme is ideal for parallel processing since different image segments can be processed at the same time. In a typical implementation, we use 10-100 loops for reaching a converged solution.

### 3.4 Field sampling and other considerations

There are several considerations for the reported FFP approach. First, we do not need to use all image segments in our implementation. Where to choose these image segments is an important consideration. In Fig. 7, we discuss different sampling strategy[34] for choosing the image segments: 1) uniform sampling over the entire FOV, 2) a higher sampling density at the center, and 3) a higher sampling density at the edge of the FOV. The ground truth aberration is generated using 320 image segments with 256 by 256 pixels each and with 25 incident angles. To test the three cases listed above, we use 81 segments with 64 by 64 pixels each and with 15 incident angles. We quantify the results via root mean squared error (RMSE). We can see that a higher sampling density at the center give a higher convergence speed.

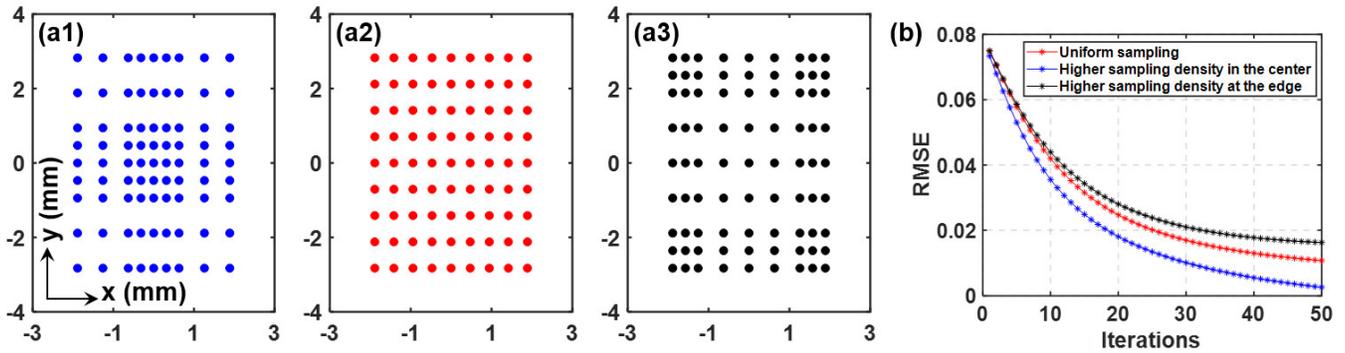

Fig. 7 Three different sampling strategies for choosing the image segments across the entire FOV. (a1)-(a3) The sampling pattern across the FOV. Each dot represents an image segment of 64 by 64 pixels. (b) The RMSE as a function of the iteration nubmer. A higher sampling density at the center enables a faster convergence speed.

Other considerations for the reported approach include the number of image segments, the number of pixels for each segment, and the number of incident angle for each segment. In our implementation, we use 81 image segments, 64 by 64 pixels per segment, and 15 incident angles. This choice is a good compromise between the convergence speed and processing time per iteration. In particular, the acquisition time of the FFP approach is less than 1s (we acquire ~20 brightfield images in total). The spatially varying aberration of the entire FOV can be recovered in ~35s using the alternating projection scheme with parallel processing of an Intel i7 CPU.

### 4. Field-dependent aberration metrology and full-field FP imaging

In this section, we demonstrate the use of the FFP approach for aberration metrology and full-field FP imaging. In our imaging setup, we illuminate the object using an LED array and acquire images using a Nikon microscope with a 2X, 0.1-NA objective lens. The object is a low-cost blood smear ($6, Carolina, human blood film slide), which has rich features over the entire FOV.

In the first experiment, we use power series (Taylor) polynomials in Fig. 3 as the aberration modes. Figure 8 shows the recovered full-field functions for 7 aberration modes (out of 17 modes in total). Similarly, we can also use Zernike polynomials in Fig. 4 as the aberration modes. The recovered full-field functions for Zernike modes are shown in Fig. 9. In this case, we have not considered the relationship between different full-field coefficients in Fig. 4. Therefore, we recover all 26 parameters listed in Fig. 4 and assume they are independent with each other.

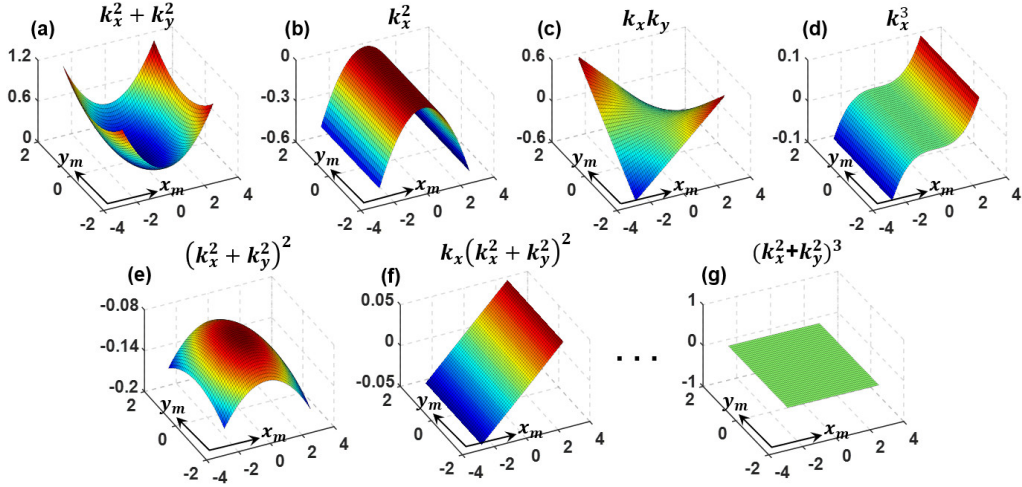

Fig. 8 Spatially varying aberrations of the microscopy imaging system with a 2X 0.1 NA objective lens. (a)-(g) The recovered field-dependent functions for 7 Taylor-polynomial aberration modes, where $(x_m, y_m)$ represents the field coordinates in millimeter and the z axis represents the amount of the aberration mode.

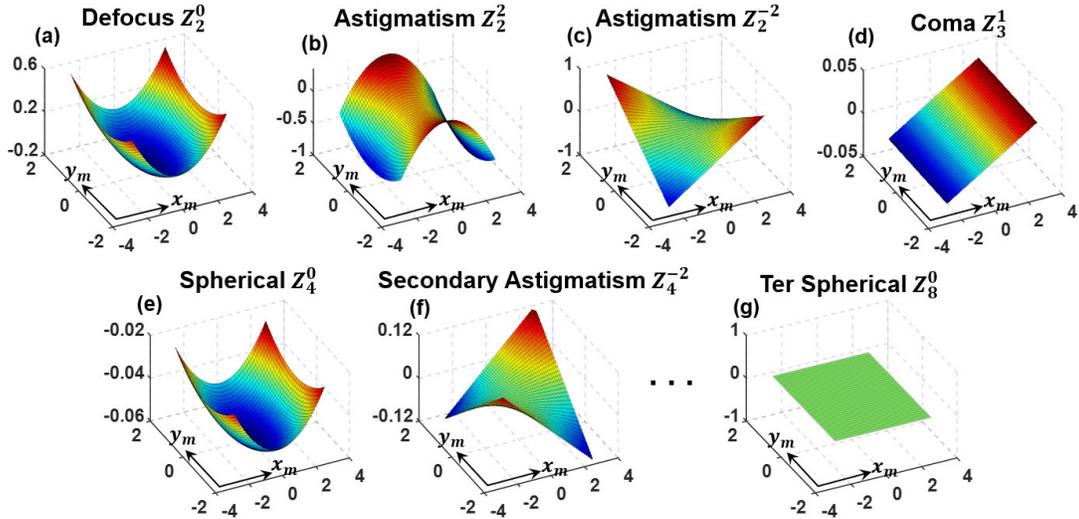

Fig. 9 The recovered field-dependent functions for 7 Zernike aberration modes, where $(x_m, y_m)$ represents the field coordinates in millimeter and the z axis represents the amount of the aberration mode.

In a regular FP implementation, one need to recover both the object and the pupil aberration for each segment. Figure 10 shows the comparison between the regular FP implementations and the reported FFP implementations. Figure 10(a) shows the raw image segment of the blood smear. We choose a region close to the edge of the FOV in this demonstration. Figure 10(b1) and 10(b2) show the recovered intensity and phase using the EPRY-FPM approach[19, 33]. Figure 10(c) shows the recovered results using the newly reported 'rPIE+Momentum' approach[35]. For both approaches, the recoveries cannot converge to a good solution due to the large pupil aberration presented in the system. To address this problem, a conventional FP implementation needs to recover the segments from the center to the edge sequentially. The recovered pupil from the central segment will be used as the initial guess for the adjacent segments. In the reported FFP, we can first recover the full-field functions as shown in Figs. 8-9. We can then obtain the pupil aberration for each segment. Finally, we can recover the high-resolution object without updating the pupil function. Different image segments can be processed in parallel since sequential pupil updating is not needed. We also note that, both the gradient-descent and the alternating projection schemes can reach the converged solution in Fig. 10(d) and 10(e). The alternating projection scheme is, in general, preferred because it allows efficient parallel processing of different image segments at the same time.

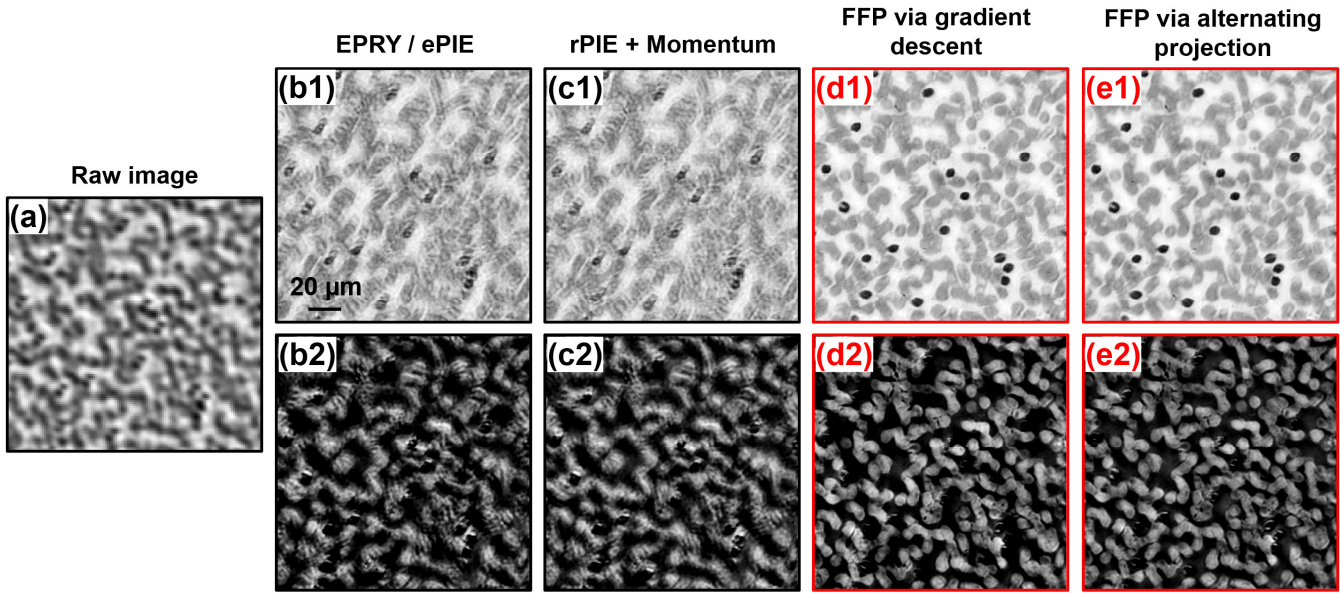

Fig. 10 Comparison between the regular FP implementations and the reported FFP implementations. (a) The raw captured low-resolution image. (b) The reconstructed object intensity and phase using the EPRY algorithm. (c) The reconstructed object intensity and phase using the 'rPIE+Momentum' approach. (d) The reconstructed object intensity and phase based on the recovered pupil from the gradient-descent-based FFP. (e) The reconstructed object intensity and phase based on the recovered pupil from the alternating-projection-based FFP.

Based on the pupil recovered from the FFP approach, we also generate a full FOV, high-resolution image of the blood smear sample in Fig. 11. In this experiment, the entire area is divided into ~320 segments. We first recover the field-dependent pupil function for each segment via the FFP approach. We then recover the object without updating the pupil function in the reconstruction process. Figure 11(a) shows the recovered full-FOV gigapixel image of the blood smear, with a synthetic NA of 0.5. The insets in Fig. 11(a) show the wavefront aberration of three regions. The aberration at the edge is much more severe compared to that at the center. Figure 11(b)-(d) show the magnified views of recovered intensity, recovered phase, and the raw images in three locations, where we can see significant resolution improvement and clear phase profiles of the blood cells.

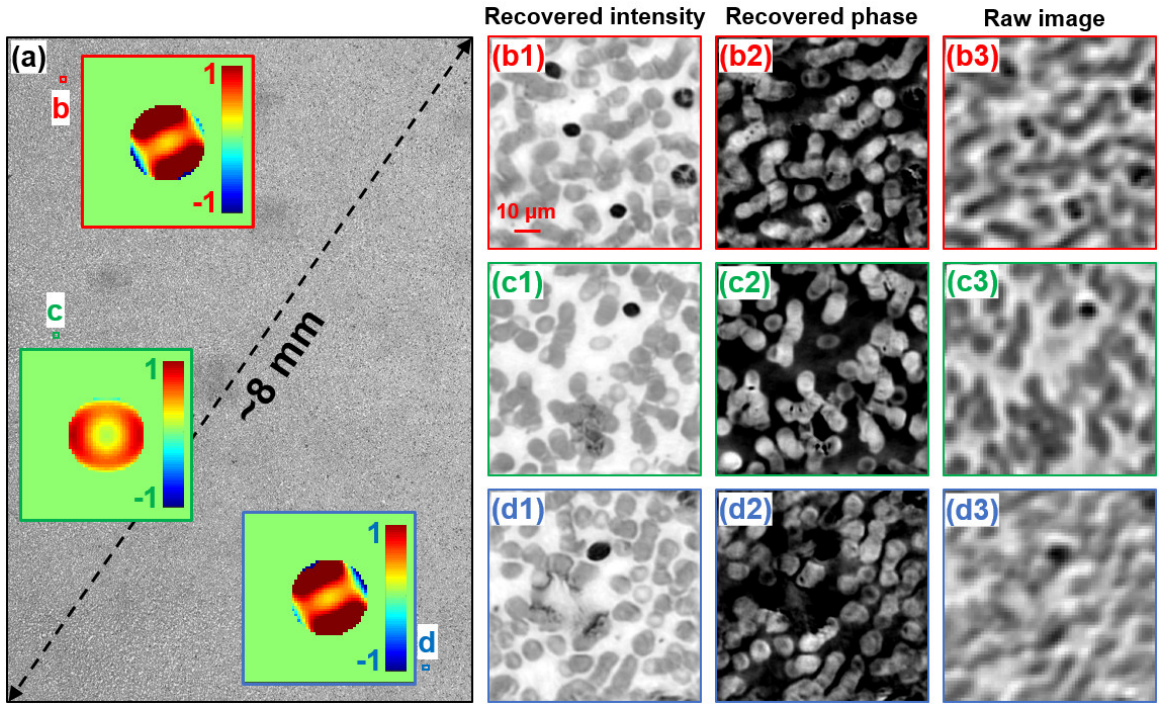

Fig. 11 Full-FOV, high-resolution reconstruction based on the pupil recovered by the FFP approach. (a) The recovered full-field intensity image of the blood smear section. Insets show the recovered pupils at three different field positions. (b1)-(d1) The magnified view of the reconstructed images at the three positions. (b2)-(d2) The recovered phase images. (b3)-(d3) The corresponding raw images.

Finally, we quantify the accuracy of the recovered pupil aberration of the reported FFP approach. In this experiment, we introduce known defocus aberrations by setting the sample to different defocus positions: z = 0 μm, 1 μm, 2 μm, 5 μm, 10 μm, 15 μm, and 20 μm. For each position, we acquire an FFP dataset and recover the full-field functions. Figure 12(a) shows two recovered full-field functions for the defocus aberration mode (i.e., the $k_x^2 + k_y^2$ term in the pupil plane). The bottom and top surfaces in Fig. 12(a) correspond to the cases by placing the object at the 0-μm and 20-μm z-positions, respectively. As such, we have the recovered pupils at both the 0-μm and 20-μm positions.

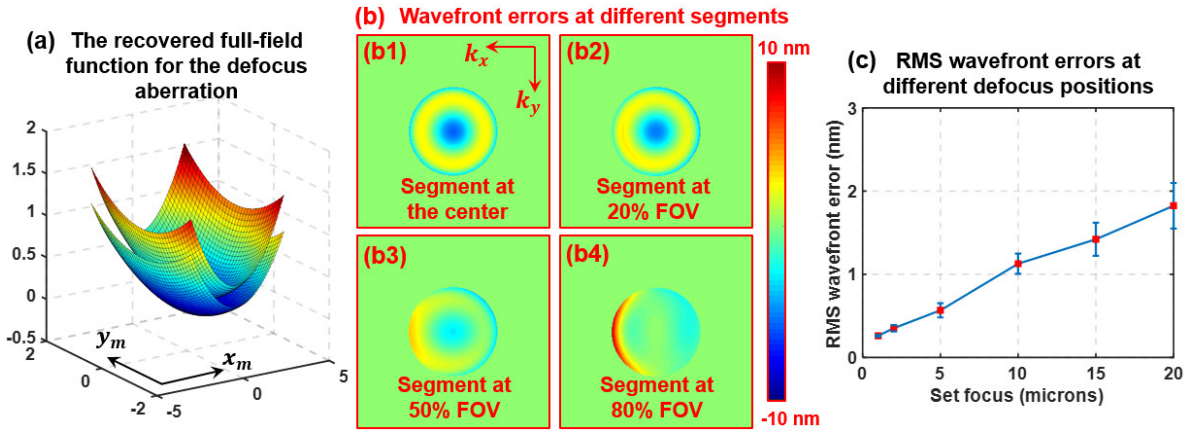

Fig. 12 Quantifying the accuracy of the recovered pupil aberrations. (a) Two recovered full-field functions for the defocus aberration mode. The bottom and top surfaces correspond to the cases by placing the object at the 0-μm and 20-μm z-positions, respectively. (b) The residue wavefront errors of recovered pupil aberrations. (c) The RMS wavefront errors of different defocus distances.

To quantify the accuracy of the recovered pupil aberrations, we first generate a standard pupil wavefront with 20-μm defocus distance. We then add this generated 20-μm defocus pupil to the recovered 0-μm pupil. The result is served as the ground-truth for 20-μm defocus pupil. The difference between the ground-truth and the recovered 20-μm pupil is the wavefront error in the reconstruction process. This wavefront error can be quantified at the different regions of the entire FOV, as shown in Fig. 12(b1)-12(b4). We can see that the wavefront errors are on the nanometer scale and there is no significant difference between different regions. We further quantify the root-mean-square (RMS) wavefront errors with different defocus distances in Fig. 12(c), where the red dots represent the average wavefront errors in all regions and the error bars represent the standard deviation at different regions of the entire FOV. Again, the RMS wavefront errors are on the nanometer scale, validating the accuracy of the recovered pupils.

## 5. Summary

In summary, we report a full-field Fourier ptychography (FFP) scheme for rapid and robust aberration metrology. The power series aberration modes and their field-dependent functions are given for modelling the spatially varying aberration in our optimization scheme. Compared to the regular aberration recovery approaches, we recover the 'full-field' coefficients instead of the pupil wavefront at each small segment. As such, the optimization degrees of freedom are orders of magnitude lower than the previous implementations. We show that the image acquisition process of FFP can be completed in ~1s and the spatially varying aberration of the entire FOV can be recovered in ~35s using a regular CPU. We also demonstrate the reported approach for generating a gigapixel image of a biological sample with a ~8-mm FOV and a 0.5 synthetic numerical aperture. The full-field functions employed in our scheme can be viewed as a strong constraint for the pupil recovery process. It can accelerate the optimization process and improve the robustness of the reconstruction. Future directions of the reported approach include the testing of other advanced optimization schemes for reducing the number of acquired images. Our on-going effort is to implement the full-field model for conventional blind deconvolution problems.

**ACKNOWLEDGMENTS**

This work was in part supported by NSF 1510077 and NIH R03EB022144.